
\input phyzzx
\mathsurround=2pt
\openup 2pt
\hfuzz 30pt
\pubnum{UTS-DFT-92-4}
\pubtype{HE}
\titlepage
\singlespace
\def\d{\partial}

\def\jcl{{\cal J}_{\rm cl}}
\def\kcl{{\cal K}_{\rm cl}}

\def\X{\dot X}
\def\r{{1\over l_{\rm m}^3}}
\def\leff{L_{\rm eff}}
\title{MEMBRANE PREGEOMETRY and the VANISHING of the COSMOLOGICAL
CONSTANT }
\author{A.Aurilia\foot{E-mail address: AAURILIA@CSUPOMONA.EDU}}
\address{Department of Physics\break
California State Polytechnic University\break Pomona, CA 91768}
\author{A.Smailagic}
\address{International Center for Theoretical Physics, Trieste, Italy\break
Strada Costiera 11-34014, Trieste, Italy}
\andauthor{E.Spallucci\foot{E-mail address: SPALLUCCI@TRIESTE.INFN.IT}}
\address{ Dipartimento di Fisica Teorica\break Universit\`a di
Trieste\break
I.N.F.N., Sezione di Trieste\break
Trieste, Italy}
\submit{ Class. Quantum Gravity }
\doublespace
\endpage
\abstract

We suggest a model of induced gravity in which the fundamental object is
a relativistic {\it membrane} minimally coupled to a background metric and
to an external three index gauge potential. We compute the low energy limit
of the two-loop effective action as a power expansion in the surface
tension.
A generalized bootstrap hypothesis is made in order to identify the physical
metric
and gauge field with the lowest order terms in the expansion of the vacuum
average of the composite operators conjugate to the background fields.
We find that the large distance behaviour of these classical fields is
described by the Einstein action with a cosmological term plus
a Maxwell type action for the gauge potential. The Maxwell term enables us
to apply the
Hawking-Baum argument to show that the physical cosmological constant
is ``~probably~'' zero.

\vfill\eject

\REFS\Aa{%
Ya.B.Zel'dovich JETP Lett. {\bf 6}, (1967), 316\hfill\break
A.D.Sakharov Sov.Phys. Dokl. {\bf 12}, (1968), 1040
}%
\REFSCON\Ab{%
S.L.Adler Rev.Mod.Phys. {\bf 54}, (1982), 729, and references therein
}%
\REFSCON\Ac{%
N.N.Khuri, Phys.Rev.{\bf D26}, (1982) 2664\hfill\break
F.David, Phys.Let.{\bf 138B}, (1984) 383
}%
\REFSCON\Ad{%
A.Aurilia, E.Spallucci ``~The role of extended objects in particle physics
and cosmology~'' in: Proc. Trieste Conf. on Super-membranes and physics in
2+1 dimensions (Trieste, July 1988).
}%
\REFSCON\Ae{%
A.Aurilia, R.S.Kissack, R.Mann, E.Spallucci Phys.Rev.{\bf D10} (1987) 2961;
\hfill\break
A.Aurilia, M.Palmer, E.Spallucci Phys.Rev.{\bf D40} (1989) 2511
}%
\REFSCON\Af{%
E.S.Fradkin, A.A.Tseytlin Nucl.Phys.{\bf B261} (1985) 1
}%
\REFSCON\Ag{%
R.D.Ball Phys.Rep.{\bf 182} (1982) 1
}%
\REFSCON\Ap{%
R.Floreanini, R.Percacci, E.Spallucci Class.Quantum Grav.{\bf 8} (1991) L193
\hfill\break
R.Floreanini, R.Percacci ``~Mean-field quantum gravity~'' INFN/AE-91/08
}%
\REFSCON\Ah{%
E.Baum Phys.Lett.{\bf 133B} (1983) 185
}%
\REFSCON\Aw{%
S.W.Hawking Phys.Lett.{\bf 134B} (1984) 403
}%
\REFSCON\Ak{%
J.B.Hartle, S.W.Hawking Phys.Rev.{\bf D28} (1983) 2960
}%
\REFSCON\Al{%
S.Coleman Nucl.Phys.{\bf B310} (1988) 643
}%
\REFSCON\Am{%
T.Banks Phys.Rev.Letters {\bf 52} (1984) 1461
}%
\REFSCON\An{%
E.Witten, in: Proceedings of the 1983 Shelter Island Conference on Quantum
Field Theory and Fundamental Problems of Physics,\hfill\break
Eds.R.Jackiw, N.Khuri, S.Weinberg and E.Witten (MIT Press, 1985)
}%
\REFSCON\Ao{%
J.D.Brown, C.Teitelboim Phys.Lett.{\bf 195B} (1987) 177
}%
\REFSCON\Aq{%
M.J.Duncan, L.G.Jensen Nucl.Phys.{\bf B336} (1990) 100
}%
\refsend
The {\it induced gravity} programme, pioneered by Zel'dovich and Sakharov in
the late sixties, provides an ingenious way to ``~sweep under the carpet~'' the
long standing problem of quantizing General Relativity\refmark{\Aa}. The
basic idea
is to consider General Relativity as an {\it effective theory}
describing the large distance spacetime structure, rather than a
fundamental theory of gravity at every length scale. The analogy which
comes
immediately to mind is with the non-renormalizable Fermi theory of weak
interactions which represents only the low-energy approximation of the
$SU(2)_L\otimes U(1)_Y$ electro-weak, renormalizable, gauge theory. In
the same spirit one would like to derive the Einstein theory as a
macroscopic
limit of some suitable gauge theory which unifies gravity with other gauge
interactions and provides a reliable description of short distance
gravitational phenomena\refmark{\Ab}.

As a by-product, this approach would provide finite, unambiguous
values of the macroscopic gravitational constants, i.e. the Newton  and the
cosmological constants, in terms of the gauge charges and vacuum
condensates
of the underlying fundamental theory.

However, there are, at least, two basic problems with this idea :

i) the fundamental field theory
one starts with is of the gauge type, with no dimensional coupling
constants, in order to implement renormalizability
at the quantum level. But
then, the induced Newton constant should arise through some dynamical
mechanism
breaking the original gauge, or Weyl, symmetry at large distances. This kind
of
process is essentially {\it non-perturbative} and is presumably similar to
the
hadronization process of the fundamental QCD degrees of freedom. Any
attempt to describe this type of phenomena, beside technical problems,
introduces regularization ambiguities which spoil the
predictability of the induced Newton constant\refmark{\Ac}.

ii) The value of the induced cosmological constant should be very small,
and possibly vanishing, in order to be consistent with the experimental
bound $\lambda_{\rm exp}\le 10^{-120}(\hbox{Planck Mass})^4$.
On the contrary, the typical value one usually
obtains is of the order $(\hbox{Planck Mass})^4$!

The current attitude towards ultra-short distance physics
is to replace local fields with extended objects, mainly
strings, as fundamental constituents of matter and to treat particle physics
below some (string)energy-scale as a local limit of the fundamental theory.
Extended objects, or p-branes,
carry a proper mass, or length scale, related to the (hyper-)
tension $\rho$ by the relation $l_{\rm p-brane}\sim (\rho)^{-{1/(p+1)}}$.
Moreover, the spatial extension of the object should improve
its ultraviolet behaviour, leading ultimately to a finite or at least
renormalizable quantum theory. Thus, one hopes that this is a good
framework
to derive unambiguously the dimensional coupling constants of the low
energy
effective theory.

In part because of the above considerations and in part because of the
increasing
relevance of relativistic membranes both in particle physics and
cosmology\refmark{\Ad}, it seems pertinent to ask if and how General
Relativity may arise as the low energy limit of a quantum theory of
relativistic membranes.  Our objective is to show that this is indeed the
case: the gravitational and gauge forces acting on the membrane are
generated by the membrane itself and the macroscopic dynamics of the
classical fields is {\it self-consistently induced} by the quantum dynamics
of the membrane in the long-wavelength approximation. The large distance
behaviour of these classical fields is described by the Einstein action with
a cosmological term plus a Maxwell type action for the three index gauge
potential, and the presence of this latter term enables us to show that the
physical cosmological constant most likely is zero.
Our starting point is the (euclidean) Nambu-Goto action for a relativistic
closed membrane interacting  with two {\it background
fields:} a symmetric, non-degenerate, tensor $J_{\mu\nu}(X)$,
and a totally antisymmetric tensor\foot{In string theories
the {\it graviton} and the {\it Kalb-Ramond gauge potential} are
present in the string spectrum.
Whether the membrane spectrum contains massless
states at all is an open (~model-dependent~) question we shall not address
here. For our purposes, we can look at equation (1) as a
generalization of the
the generally covariant action of a point-particle coupled to an external
electromagnetic field.
}
$K_{\mu\nu\rho}(X)$
$$\eqalign{
&S_{\rm NG}=\r\Bigl[\int_{\cal H}d^3\sigma\sqrt{{1\over 3!}
J_{\mu\mu'}J_{\nu\nu'}J_{\rho\rho'}\X^{\mu\nu\rho}
\X^{\mu'\nu'\rho'}}+
{1\over 3!}\int_{\cal H} d^3\sigma \X^{\mu\nu\rho} K_{\mu\nu\rho}(X)\Bigr]
\ ,\cr
&\X^{\mu\nu\rho}\equiv\delta^{[abc]} \partial_a X^\mu\partial_b X^\nu
\partial_c X^\rho\ ,\quad J_{\mu\nu}=J_{\mu\nu}(X)\ ,\cr}
\eqno(1)
$$
where ${\cal H}$ stands for a domain in the space of the parameters
$\sigma^a=(\sigma^1,
\sigma^2,\sigma^3)$ which represents the euclidean membrane manifold and
$\X^{\mu\nu\rho}$ stands for the tangent three-vector at each point of the
embedded
submanifold $x^\mu=X^\mu(\sigma)$ which represents the world-history of
the
membrane in the (euclidean) spacetime. Finally,
for later convenience, we have expressed the gauge coupling constant in
terms of the {\it surface tension} $1/l_{\rm m}^3$,
and rescaled the gauge field $K$
so that it becomes adimensional, $[K]=1$.

We remark the absence of any kinetic term for $J_{\mu\nu}$ and
$K_{\mu\nu\rho}$.
Our final goal is just to recover these terms from the quantum dynamics
of the membrane.

Usually, this action is interpreted as describing the classical dynamics of
the extended object under the combined effects of the external gauge
potential
$K_{\mu\nu\rho}$ and the pre-assigned gravitational field $J_{\mu\nu}$.
Thus, it
corresponds to the limit of Classical Bubble Dynamics\refmark{\Ae}
where both gauge and
gravitational degrees of freedom are frozen. However, at this
early stage, the tensors $J_{\mu\nu}$ and
$K_{\mu\nu\rho}$ play simply the role of {\it auxiliary field variables}
introduced to endow the model with:

\noindent
i) {\it general covariance} in
``~target space~'', and,

\noindent
ii) extended gauge invariance  $\displaystyle{K_{\mu\nu\rho}\rightarrow
K_{\mu\nu\rho}+l_{\rm m}\d_{\,[\mu}\Lambda_{\nu\rho]}}$.

\noindent
Indeed these two
requirements alone are sufficient to determine the form of the effective
lagrangian which, in turn, will provide a physical interpretation of
$J_{\mu\nu}$ and $K_{\mu\nu\rho}$.
Variation of the action with respect to the
external fields gives we the corresponding ``~current densities~'':
$$\eqalignno{
&{\delta S_{\rm NG}\over\delta J_{\mu\nu}}\equiv
{1\over 2}T^{\mu\nu}(X)
={1\over l_{\rm m}^3} \sqrt\gamma\gamma^{ab}\d_a X^\mu\d_b X^\nu
&(2a)\cr
&{\delta S_{\rm NG}\over\delta K_{\mu\nu\rho}}=\r\X^{\mu\nu\rho}
=\r\delta^{[abc]}\partial_aX^\mu\partial_bX^\nu\partial_cX^\nu\ ,
&(2b)\cr
}
$$
where $T^{\mu\nu}(X)$ is the membrane energy-momentum tensor-density,
$\gamma_{ab}$ is the induced
metric on the membrane world-tube, i.e.
$\gamma_{ab}=\d_a X^\mu\d_b X^\nu J_{\mu\nu}$,
and the ``~gauge-current~'' is represented by the tangent three-vector
$\X^{\mu\nu}$.
Therefore, $J_{\mu\nu}$ and $K_{\mu\nu\rho}$ can also be considered as
{\it external
sources} for the composite operators
$T^{\mu\nu}(X)$ and $\X^{\mu\nu}$.

{}From now on, we shall make no distinction between
background fields and external sources.

The generating functional corresponding to the classical action (1) is
$$
Z[J_{\mu\nu},K_{\mu\nu\rho}]=\sum_{\rm topologies}
\int\prod_\xi\sqrt{{\rm det}J_{\mu\nu}(X)}
{\cal D}[X]\exp\left[-S_{\rm NG}\right]\ ,
\eqno(2)
$$
where the jacobian $\sqrt{{\rm det}J_{\mu\nu}(X)}$ has been inserted into
the
functional measure to preserve general covariance
in the target space;
gauge fixing and ghost terms required for
the reparametrization invariance of the world-track are understood but
were purposely omitted in order to avoid an unnecessarily complicated
formalism.

{}From our vantage point, the most relevant property of the partition function
of the
quantum membrane is that an {\it integration over spacetime points is
implicit in the functional integral}.
In fact, the free membrane term is insensitive
to the position of the membrane ``~centre~'', that is, the action (1) is
invariant under the transformation
$\displaystyle{X^\mu(\sigma)\rightarrow
X^\mu(\sigma)+x^\mu}$ where
$x^\mu$ is constant, i.e., independent of the world-tube coordinates
$\sigma$.
Hence, the free membrane partition
function contains the zero-mode contribution (the spacetime volume) as a
factor.
But, in the presence of background fields, translational invariance is broken
and the four dimensional zero-mode integral becomes non-trivial\refmark{\Af}.
Then, we find it useful
to extract the, integration over the membrane ``~centre~'' from the very
beginning by
separating the zero mode contribution as follows,
$\displaystyle{X^\mu(\sigma)=x^\mu+\eta^\mu(\sigma)}$; then,
$$\eqalign{
\int {\cal D}[X]F[X]&=\int d^4x\int {\cal D}[\eta]F[x+\eta] \ ,\cr
{\cal D}[X]&=D[\eta]\delta^4(P^\mu[x,\eta])\Delta_{\rm FP}[x,\eta]\ ,\cr
\Delta_{\rm FP}&={\rm Det}
\left({\d P^\mu[a+\eta]\over\d a^\nu}\right)_{x=0}\ .\cr}
\eqno(3)
$$
Here $P^\mu[x,\eta]=0$ is the gauge fixing condition breaking the invariance
under $\eta\rightarrow\eta+{\rm const.}$, and $\Delta_{\rm FP}[x,\eta]$ is
the corresponding ghost determinant. With the aid of equation (3) we can
write $Z$ as
$$\eqalign{
Z&=\int d^4x\sqrt{J(x)}\leff[J,K]\cr
\leff[J,K]&=\sum_{\rm topologies}\int D[\eta]
\exp\left[-S(x+\eta, K)\right]\ .\cr}
\eqno(4)
$$
In principle, $\leff$ depends on all powers of the derivatives of the fields
multiplied
by suitable powers of $l_{\rm m}$. However, the gauge simmetries of the
model
force the derivatives of $J_{\mu\nu}(x)$ and $K_{\mu\nu\rho}(x)$
to appear only through the curvature tensor of the
$J_{\mu\nu}$-metric and the field strength
$K_{\mu\nu\rho\sigma}=\d_{\,[\mu}K_{\nu\rho\sigma]}$ .
The $\sqrt{J(x)}$ factor comes from the zero-mode contribution to
the functional covariant measure in equation (2).

On general grounds, $\leff$ will be a non-local quantity difficult to
compute.
However, our
purpose is to determine the ``~{\it low energy}~'' approximation of $\leff$,
i.e., the leading terms in the $l_{\rm m}$ power expansion\foot{ This
approximation is similar to the inverse mass power expansion of the
effective action in chiral gauge theories\refmark{\Ag}. The mass of the
field
is here replaced by the membrane tension.}, which, at some energy scale
below the Planck energy, dominate the {\it local} part of the effective
action. Furthermore we do not consider branching and rejoining processes
and take into account only the contribution to the functional integral
coming from ``~free~'' membranes emerging from a point and finally recollapsing
to a point.
In this approximation we interpret equation (2)
as the generating functional for the
vacuum-to-vacuum amplitude in the presence of the background
fields $J$ and $K$ and the general form of $\leff$ is determined by the
requirements of general covariance and gauge symmetry alone :
$$\eqalign{
Z[J,K]&={1\over l_{\rm m}^4}
\int d^4x\sqrt J\Bigl[2\Lambda\cr
&+l_{\rm m}^2\left(a\,J^{\mu\nu}R_{\mu\nu}(J)-b\,K_{\mu\nu\rho\sigma}
K_{\mu'\nu'\rho'\sigma'}J^{\mu\mu'}J^{\nu\nu'}J^{\rho\rho'}J^{\sigma\sigma'
}
\right)+0(l_{\rm m}^4)\Bigr]\cr}\ ,
\eqno(6)
$$
in terms of three positive numerical constants $a$, $b$, and $\Lambda$ .
It is worth noticing that similar approach has been adopted for string as well
\refmark{\Af}, but in that case the abovementioned constants
have an unpleasant
cut-off dependence. In this case however, it turns out that the constants
are {\it finite} which makes membranes more appealing pregeometric objects.

Variation of $Z[J,K]$ with respect to the  external
sources provides now the corresponding vacuum average of the composite
field operators coupled to them, i.e. the so called {\it classical fields}:
$$\eqalignno{
&\jcl^{\mu\nu}\left(J_{\rho\sigma};K_{\rho\sigma\tau}\right)
\equiv{\delta Z\over\delta J_{\mu\nu}(x)}   &(7a)\cr
&\kcl^{\mu\nu\rho}\left(J_{\rho\sigma};K_{\rho\sigma\tau}\right)
\equiv{\delta Z\over\delta K_{\mu\nu\rho}}\ . &(7b)\cr
}
$$
Thus, an {\it effective dynamics} for the background fields emerges at the
quantum level. In fact, it is   customary to derive the
``~effective action~'' for the classical fields
by exchanging the external sources in favour of the
classical fields by means of the functional Legendre transform
$$
\Gamma^{\rm eff}\left(\jcl^{\mu\nu},\kcl^{\mu\nu\rho}\right)=
\int d^4x\left[\jcl^{\mu\nu}J_{\mu\nu}+\kcl^{\mu\nu\rho}K_{\mu\nu\rho}
\right]
-Z\left(J_{\mu\nu}, K_{\mu\nu\rho}\right)\ ,
\eqno(8)
$$
and then deriving effective field equations for the classical fields by varying
$\Gamma^{\rm eff}$.

Before we attempt to implement this algorithm, we should perhaps
emphasize that while the background fields are put in ``~by hand~'' in the
action in order to implement some symmetry principle, the classical fields
are introduced dynamically in the theory as vacuum averaged
values of suitable composite operators. In principle these c-number
quantities can be expressed perturbatively in terms of the external sources
once $Z(J,K)$
is determined in powers of $l_{\rm m}$.

Our basic assumption, then, is the following
\underbar{\it generalized bootstrap hypothesis:}

\noindent
{\it at the lowest order in the expansion parameter, both the classical and
the background fields coincide with
the macroscopic fields
in which the membrane moves.} In other words,
over distances much larger than $l_{\rm m}$, we can no longer distinguish
between external sources, vacuum expectation values and the ``~classical
forces~'' acting on the membrane.\foot{ A similar self-consistency criterion
has been recently applied in the framework of a ``~mean field~'' quantization
of a $GL(4)$ gauge theory of gravity.\refmark{\Ap} Also in this case one has
both the soldering form as dynamical variable and an auxiliary metric
introduced into the model only to allow perturbative calculations. At the end
the vacuum expectation value of the composite operator describing the physical
metric is self-consistently identified with the background metric itself.}

{}From eq.(7) we obtain the explicit form of the classical fields in powers of
$l_{\rm m}^{-2}$
$$\eqalign{
\jcl^{\mu\nu}=&
{\Lambda\over l_{\rm m}^4}\sqrt J J^{\mu\nu}\cr
&-{1\over l_{\rm m}^2}\sqrt J
\left[a\left(R^{\mu\nu}-{1\over 2}J^{\mu\nu}R\right)+4b\left(
K^\mu{}_{\lambda\rho\sigma}K^{\nu\lambda\rho\sigma}-{1\over
8}J^{\mu\nu}
K_{\lambda\rho\sigma\tau}K^{\lambda\rho\sigma\tau}\right)\right]+\dots\cr
\kcl^{\mu\nu\rho}=&
{48b\over l_{\rm m}^2}\sqrt J\,\nabla_\lambda
K^{\lambda\mu\nu\rho}+\dots
\cr}
\eqno(10)
$$
where $\nabla_\lambda $ is the torsion free covariant derivative
compatible
with the $J_{\mu\nu}$ metric
Correspondingly, we expand the sources as
$$\eqalign{
&J_{\mu\nu}=g_{\mu\nu}+l_{\rm m}^2 j_{\mu\nu}+\dots\cr
&J^{\mu\nu}=g^{\mu\nu}-l_{\rm m}^2
g^{\mu\mu'}g^{\nu\nu'}j_{\mu'\nu'}+\dots\cr
&\sqrt J=\sqrt g\left(1+{l_{\rm m}^2\over
2}g^{\mu\nu}j_{\mu\nu}\right)\cr
&K_{\mu\nu\rho}=K^{0)}_{\mu\nu\rho}+l_{\rm m}^2
K^{1)}_{\mu\nu\rho}+\dots\cr}
\eqno(11)
$$
and, following our bootstrap hypothesis, set
$$\eqalign{
&\jcl^{\mu\nu}\equiv{\Lambda\over l_{\rm m}^4} \sqrt g g^{\mu\nu}\cr
&\kcl^{\mu\nu\rho}\equiv
{48b\over l_{\rm m}^2}\sqrt g\,\nabla_\lambda H^{\lambda\mu\nu\rho}\
,\cr
&H_{\lambda\mu\nu\rho}\equiv\partial_{\,[\lambda}B_{\mu\nu\rho]}\cr
}
\eqno(12)
$$
where the covariant derivative is now computed by means of the
{\it macroscopic metric} $g_{\mu\nu}$.
Then, we find at the $l_{\rm m}^{-2}$ order,
$$\eqalignno{
&K^{0)}_{\mu\nu\rho}=B_{\mu\nu\rho} &(13a)\cr
&\Lambda j^{\mu\nu}=-a R^{\mu\nu}(g)+
4b\left[K^{\mu\rho\sigma\tau}K^\nu{}_{\rho\sigma\tau}
-{3\over8}g^{\mu\nu}K^{\lambda\rho\sigma\tau}
K_{\lambda\rho\sigma\tau}\right]\ . &(13b)
\cr}
$$
Now, we can evaluate the Legendre transform $(8)$ and compute the
effective
action for the classical fields $g$ and $B$ up to the order $l_{\rm m}^{-2}$:
$$\eqalign{
\Gamma^{\rm eff}\left(g_{\mu\nu},B_{\mu\nu\rho}\right)&=
\int d^4x\sqrt g\left[\left(
{4\Lambda\over l_{\rm m}^4}-{a\over l_{\rm m}^2}R(g)-
{4b\over l_{\rm m}^2}H_{\mu\nu\rho\sigma}H^{\mu\nu\rho\sigma}\right)-
{2b\over l_{\rm
m}^2}H_{\mu\nu\rho\sigma}H^{\mu\nu\rho\sigma}\right]\cr
&-{1\over l_{\rm m}^4}
\int d^4x\sqrt g [2\Lambda
+l_{\rm m}^2(a\,g^{\mu\nu}R_{\mu\nu}(g)-b\,H_{\mu\nu\rho\sigma}
H^{\mu\nu\rho\sigma})\cr
&-l_{\rm m}^2\left(a\,g^{\mu\nu}R_{\mu\nu}(g)+2b\,H_{\mu\nu\rho\sigma}
H^{\mu\nu\rho\sigma}\right)
+0(l_{\rm m}^4)]\cr
&=\int d^4x\sqrt g\left[{2\Lambda\over l_{\rm m}^4}-{a\over l_{\rm m}^2}
R(g)-{b\over l_{\rm m}^2}H_{\mu\nu\rho\sigma}H^{\mu\nu\rho\sigma}+
0(1)\right]\ .\cr}
\eqno(14)
$$
If we define the induced constants as
$$
\eqalignno{
&{a\over l_{\rm m}^2}\equiv {1\over 16\pi G_{\rm N}}\ ,&(15a)\cr
&{2\Lambda\over l_{\rm m}^4}\equiv
{2\lambda_{\rm ind}\over 16\pi G_{\rm N}}\ ,&(15b)\cr}
$$
and rescale the generalized Maxwell field strength according to
$$
{b\over l_{\rm m}^2}H_{\mu\nu\rho\sigma}H^{\mu\nu\rho\sigma}\equiv
{1\over 2\cdot 4!}F_{\mu\nu\rho\sigma}F^{\mu\nu\rho\sigma}\ ,
\eqno(16)
$$
we obtain that the low energy limit of the membrane effective action is
the Einstein action with a cosmological term, coupled to the $F$-field:
$$
\Gamma^{\rm eff}\left(\jcl^{\mu\nu},\kcl^{\mu\nu\rho}\right)=
-\int d^4x\sqrt g
\left[{1\over 16\pi G_{\rm N}}\left(R(g)-2\lambda_{\rm ind}\right)
+{1\over 2\cdot 4!}F_{\mu\nu\rho\sigma}F^{\mu\nu\rho\sigma}\right]\ .
 \eqno(17)
$$
In order to reproduce the correct value of $G_{\rm N}$ the constant $l_{\rm
 m}$ must
 be of the order of the Planck length, i.e. $l_{\rm m}\sim L_{\rm P}$,
and our pregeometric membranes have surface tension
$\sim (\hbox{Planck Mass})^3$; correspondingly, the
induced Cosmological Constant turns out to be, as usual, enormously large:
$\lambda_{\rm ind}\sim L_{\rm P}^2$. However, this result is not as bad as
it
would seem at first sight because $\lambda_{\rm ind}$ {\it is not the
physical
cosmological constant !} In fact, the generalized Maxwell field strength,
in four dimensions, has no propagating modes associated to it, rather it
represents a constant energy density background which shifts the value of
the
cosmological constant. The Maxwell equation derived from
eq.(17) admits a solution of the form
$$
\bar F_{\mu\nu\rho\sigma}=m^2\delta_{[\mu\nu\rho\sigma]}\ ,
\quad m=\hbox{const}\ ,\quad [m]=\hbox{mass}\ .\eqno(18)
$$
When inserted back in $\Gamma^{\rm eff}$, the solution (18) gives the
Einstein action with a new effective cosmological constant, $\lambda_{\rm
phys}$,
given by
$$
\lambda_{\rm phys}(m)=\lambda_{\rm ind}-8\pi G_{\rm N}m^4\ .
\eqno(19)
$$
Thus, the physical spacetime emerging from the underlying quantum
dynamics
is a solution of the vacuum Einstein equations with a cosmological term
$$
R_{\mu\nu}-{1\over 2}g_{\mu\nu}\left(R-2\lambda_{\rm phys}(m)\right)=0
\eqno(20)
$$
The actual
value of $\lambda_{\rm phys}(m)$ can now be fixed by the Hawking-Baum
argument\refmark{\Ah,\Aw}. Supplemented with the Hartle-Hawking
boundary
condition\refmark{\Ak} the effective Einstein equation (20) admits as
solution the 4-sphere of radius
$r=\left((3/\lambda_{\rm phys}(m)\right)^{1/2}$
to which correspond an action
$$
\Gamma^{\rm eff}(S^{4)})=-{3\pi\over G_{\rm N}\lambda_{\rm phys}(m)}\ .
\eqno(21)
$$
If one interprets $\displaystyle{\exp \left(-\Gamma^{\rm eff}\right)}$
as a probability
distribution for the value of the physical cosmological
constant, then the peak for $\lambda_{\rm phys}(m)\rightarrow 0^+$
suggests
that the most probable value of the physical cosmological constant is zero.
The key property which allows us to apply the Hawking-Baum argument is
the
general property that (p+1) classical gauge forms in a (p+2) dimensional
spacetime
describe a constant background energy distribution rather than propagating
degrees of freedom\refmark{\Ad}. In particular, the possibility that four-
forms may be used to circumvent the problem of the cosmological constant,
either in conjunction with the Baum-Hawking mechanism or with Coleman's
mechanism\refmark{\Al}, has been suggested by several
authors\refmark{\Am,\An,\Ao,\Aq}. To our knowledge, however, the model
discussed in this paper is the first one in which the Maxwell term for a
four-form arises naturally in the low energy effective lagrangian for {\it
induced} gravity; our discussion makes it clear that the presence of this
term can be traced back to the use of a relativistic membrane, rather than a
local field, or even a string, as the basic pregeometric object in the action
$(1)$.

\refout
\bye